\documentclass[fleqn,usenatbib,useAMS]{aastex631}
\usepackage{graphicx}
\usepackage{bm}
\usepackage{amssymb}
\usepackage{amsmath}
\usepackage{amsfonts}
\usepackage{mathrsfs}
\usepackage{epsfig}
\usepackage{mathtools}
\usepackage{enumerate}
\usepackage{color}

\usepackage{subcaption}

\shorttitle{Spherical accretion in Newtonian analogous construct}
\shortauthors{S. Ghosh et al.}

\begin{document}


\title{Spherical accretion in the Schwarzschild spacetime in the Newtonian analogous construct} 


\author{Shubhrangshu Ghosh}
\affiliation{Center for Astrophysics, Gravitation and Cosmology, Shri Ramasamy Memorial (SRM) University Sikkim, \\
5$^{\rm th}$ Mile Tadong, Gangtok, 737102, India} 
\affiliation{Department of Physics, Shri Ramasamy Memorial (SRM) University Sikkim, 5$^{\rm th}$ Mile Tadong, Gangtok, 737102, India} 
\email{shubhrangshughosh.r@srmus.edu.in} 

\author{Souvik Ghose}
\affiliation{Harish-Chandra Research Institute, HBNI, Chhatnag Road, Jhunsi, Allahabad, 211109, India} 
\email{souvikghose@hri.res.in}

\author{Kalyanbrata Pal}
\affiliation{Harish-Chandra Research Institute, HBNI, Chhatnag Road, Jhunsi, Allahabad, 211109, India} 
\email{kalyanbratapal@hri.res.in}

\author{Arunabha Bhadra}
\affiliation{High Energy \& Cosmic Ray Research Centre, University of North Bengal, Siliguri, 734013, India } 
\email{aru\_bhadra@yahoo.com}

\author{Tapas K. Das}
\affiliation{Harish-Chandra Research Institute, HBNI, Chhatnag Road, Jhunsi, Allahabad, 211109, India}
\affiliation{Physics and Applied Mathematics Unit, Indian Statistical Institute, 203 Barrackpore Trunk Road, Kolkata, 700108, India}
\email{tapas@hri.res.in}

\correspondingauthor{Tapas K. Das}
\email{tapas@hri.res.in}



\begin{abstract}

The velocity-dependent Newtonian analogous potentials (NAPs) corresponding to general relativistic (GR)
spacetimes accurately capture most of the relativistic features, including all classical tests of GR, effectively
representing spacetime geometries in Newtonian terms. The NAP formulated by Tejeda \& Rosswog (TR13) for
Schwarzschild spacetime has been applied to the standard thin accretion disk around a black hole (BH) as well as
in the context of streamlines of noninteracting particles accreting onto a Schwarzschild BH, showing good
agreement with the exact relativistic solutions. As a further application, here we explore the extent to which TR13
NAP could describe a transonic hydrodynamical spherical accretion flow in Schwarzschild spacetime within the
framework of standard Newtonian hydrodynamics. Instead of obtaining a typical single ``saddle-type'' sonic
transition, a ``saddle$-$spiral pair'' is produced, with the inner sonic point being an (unphysical) ``spiral type'' and the
outer being a usual ``saddle type.'' The Bondi accretion rate at outer sonic radii, however, remains consistent with
that of the GR case. The primary reason for the deviation of our findings from the classical Bondi solution is likely
due to the inconsistency between the Euler-type equation in the presence of velocity-dependent TR13 NAP within
the standard Newtonian hydrodynamics framework, and the corresponding GR Euler equation, regardless of the
fluid’s energy. Our study suggests that a (modified) hydrodynamical formalism is needed to effectively implement
such potentials in transonic accretion studies that align with the spirit of TR13 like NAP, while remaining
consistent with the GR hydrodynamics. This could then essentially circumvent GR hydrodynamics or GR
magnetohydrodynamics equations

\end{abstract}


\section{Introduction}
\label{sec:intro}

In recent times, Newtonian-like analogous potentials (hereinafter NAPs) of 
corresponding general relativistic (GR) spacetime geometries have been constructed in the literature (\citet{tejeda2013accurate}, hereinafter TR13; \citet{ghosh2014newtonian, ghosh2015newtonian,sarkar2014s, ghosh2016exact}), with an intent to have an 
accurate or correct representation of spacetime geometries (or metric theories of gravity) in the Newtonian framework. Following TR13, most of these NAPs have been derived from the 
Hamiltonian of the test particle motion, starting from the actual geodesic motion of test particles in corresponding spacetimes 
(TR13; \citet{ghosh2014newtonian, ghosh2015newtonian, sarkar2014s}). The 
analogous potentials comprise of the terms containing explicit information of the velocity of the particle motion, in addition to the 
spherically symmetric part of the gravitational source and the other source specific terms. 

In the far (weak) field gravity at (nearly) static conditions, spherically symmetric Newtonian gravitational potential continues 
to be used in almost all applications as a valid approximation, however, its description 
fails to account for most of the effects in strong field gravity, even approximately. Previously, efforts have been undertaken to 
modify the Newtonian potential by tweaking its form to mathematically replicate or mimic certain GR features of corresponding spacetime geometries, primarily 
to describe the inner relativistic accretion flow dynamics in the vicinity of black holes (BHs)/compact objects. These modified potentials are commonly referred to as 
pseudo-Newtonian potentials (PNPs). Several such PNPs have been introduced in the literature soon after its inception by 
\citet{paczynsky1980thick}(hereinafter PW80) to reproduce certain relativistic features of Schwarzschild BH, either through 
ad hoc proposition or following certain specific methods (see for e.g., \citet{mukhopadhyay2002description, ghosh2007generalized}). For an elaborate discussion regarding 
various aspects of PNPs, see for e.g., TR13; \citet{ghosh2016exact, witzany2017pseudo} and references in them. While prescribing the PNPs, the emphasis are 
mostly laid to correctly reproduce the innermost stable and bound circular orbits, without endeavoring to actually construct any Newtonian analogues of corresponding 
spacetime geometries. PNPs, thus, lack the uniqueness to describe the dynamical properties of corresponding spacetimes in its entirety, within a reasonable accuracy. 

As the effect of a spacetime geometry could only be perceived through the dynamics of geodesic motion, any 
correct analogous construct of metric theory of gravity in the Newtonian framework should  
then explicitly contain the information of the velocity of the particle motion. In this context, more recently, adhering to a 
first-principle approach, \citet{ghosh2016exact} formulated an exact relativistic Newtonian analogue 
of spherically symmetric static spacetime metrics, described through a relativistic gravitational action in the Newtonian 
picture, furnishing identical geodesic equations of motion to those of the parent metric. 
The corresponding velocity-dependent NAP exactly reproduces 
all relativistic gravitational features of corresponding spacetime geometry for a distant stationary observer, including all the existing 
classical tests of GR. In general, velocity-dependent NAPs are far more accurate representations of relativistic spacetime geometries in analogous 
Newtonian framework, with an ability to uniquely describe the corresponding spacetime dynamics more precisely 
(for more details, see the references in the first paragraph of this section). For more comments on velocity dependent NAPs and their interpretation, 
readers can follow \citet{witzany2017pseudo, friedman2019geometrization, friedscarr2019}. 

One of the potential astrophysical scenarios where modified Newtonian-like potential could be useful is the relativistic accretion phenomena 
around BHs/compact objects. In this regard, the existing PNPs, particularly PW80 or for that matter, \citet{mukhopadhyay2002description} or \citet{ghosh2007generalized} potentials 
had been used quite extensively in the astrophysical literature, to study the complex inner accretion flow dynamics around non-rotating/rotating BHs 
(e.g., \citet{matsumoto1984viscous, abramowicz1988slim,chakrabarti1995spectral,chakrabarti1996grand,hawley2002high,hawley2002dynamical,igumenshchev2003three,chan2005spectral,lipunov2007extra,igumenshchev2008magnetically,shafee2008viscous,benson2009maximum,
bhattacharya2010disk,ohsuga2011global,narayan2011bondi,yuan2014hot} and references 
therein; \citet{bu2016magnetohydrodynamic,mondal2019ultraluminous,dihingia2020properties})
The approximation scheme through which the PNPs are implemented is by simply replacing the Newtonian gravitational potential (or force) by these 
modified potentials in the Newtonian hydrodynamical equations. As velocity-dependent NAPs are more accurate representations of corresponding 
spacetime metrics, it would be then worthwhile to investigate, how far one can make use of this formalism in hydrodynamical accretion studies within the framework of standard Newtonian hydrodynamics. The NAP developed by TR13 under the low-energy limit condition, for Schwarzschild spacetime, which can accurately capture most of all the corresponding relativistic features of Schwarzschild spacetime, including the classical experimental tests of GR, has been applied to the standard geometrically thin accretion disk, and is found to be in good agreement with the relativistic value of radiative energy flux emitted from the disk. In addition, the said NAP can reproduce the relativistic streamlines of non-interacting 
particles accreting onto a Schwarzschild BH  (\citet{tejeda12accrnonint,tejeda2013acrrnonint}; TR13). \citet{bonnerot2016disc} have recently used the velocity-dependent analogous potential of TR13 effectively, in the context of tidal disruption of stars by 
Schwarzschild BHs, within the framework of smooth particle hydrodynamics (SPH).

As a further astrophysical application, we, here, explore the extent to which TR13 NAP could describe a transonic radial/quasi-radial type (hydrodynamical) accretion flow in Schwarzschild spacetime, i.e., a Bondi-type hydrodynamical spherical accretion \citet{bondi1952spherically} within the framework of standard Newtonian hydrodynamics, by performing an actual hydrodynamical 
study. 
Extensive investigations on Bondi-type spherical flows have been undertaken by several authors to address the 
critical properties of such flows, and have been revisited time and again from multiple angles in both GR as well as in pseudo-GR regimes 
(e.g., \citet{michel1972accretion, flammang1982stationary,turolla1989role,chang1985standing,nobili1988henyey,nobili1991spherical,das2001pseudo,tapas2002,mandal2007critical,ghoshbanik2015,korol2016bondi,ciotti2017isothermal,ramirez2018impetus,raychaudhuri2018spherical,raychaudhuri2021multi,ramirez2019bondi,richards2021relativistic,yang2021spherical,aguayo2021spherical,mancino2022polytropic}). Recently, few authors have 
also investigated such Bondi-type spherical accretion in modified Einsteinian gravities (e.g., \citet{john2019bondi,kalita2019asymptotically,bauer2022spherical}). Classical Bondi flow onto a point gravitating mass exhibits 
transonic behavior, usually described by a single (saddle-type) critical transition. However, instances of violation of this ``criticality-condition'' or appearance of 
multi-critical points in spherical accretion has been addressed by several authors in the literature (e.g., \citet{flammang1982stationary,turolla1988general,turolla1989role,nobili1991spherical,mandal2007critical,ghoshbanik2015,raychaudhuri2018spherical,raychaudhuri2021multi}). 
\citet{mandal2007critical}, in fact, discussed the possibility of having three critical points in the spherical accretion flow onto a 
Schwarzschild BH in the GR framework, through dynamical systems analysis 
\footnote{For more details on the astrophysical relevance of Bondi-type spherical accretion, readers can follow \citet{raychaudhuri2018spherical,raychaudhuri2021multi}, and references therein}. 


The rest of the paper is as follows: In the next section we will formulate our hydrodynamical model for Bondi-type spherical accretion in the presence of the analogous potential of TR13, 
and perform the global critical point analysis in the parameter space. In \S 3 we investigate the nature of critical points from dynamical system analysis. \S 4 deals with 
the fluid behavior of spherical type flows in the presence of such velocity-dependent 
potential and flow topology for such a system. Finally, we culminate in \S 5 with a discussion and general remarks on our findings.

\section{Hydrodynamical formalism of spherical accretion flow in the presence of velocity-dependent analogous potential and global critical point analysis}
\label{sec:hydro}

Preserving the conventional approach, we consider a quasi-stationary spherically symmetric accretion flow onto a Schwarzschild BH. Consequently, we assume the flow to 
be inviscid in nature. Here it is to be noted that although `${r\phi}$' component of viscous stress tensor is discarded owing to the consideration of spherical 
flow, however, `${rr}$' component of the stress tensor might not altogether be neglected. Nonetheless, following the arguments in \citet{narayan2011bondi} and 
\citet{raychaudhuri2018spherical}, we discard the above stated stress tensor term in our present work. It would be interesting to investigate the effect of this term on the Bondi solution, which however, is beyond the scope of the present work, and left for future analysis. As the flow is inviscid, we further neglect the heat generation and radiative loss from the system. Being spherically symmetric, we express the dynamical flow variables only as functions of 
radius $r$. Throughout our analysis, $r$ is expressed in units of $r_g = G M_{\rm BH}/c^2$, where $M_{\rm BH}$ is the BH mass, $c$ the speed of light, and $G$ the 
usual gravitational constant. Radial velocity $v$ and sound speed $c_s$ are expressed in units of $c$. The accretion flow is considered to be polytropic with equation of state 
given by $P = K \rho^{1+1/n}$, where $P$ and $\rho$ are the usual gas pressure and density of the accreting plasma. $K$ is a constant which contains the information
of the entropy content of the flow. The polytropic index $n$ is related to the adiabatic index $\gamma$ through the usual relation $n = \frac{1}{\gamma -1}$, where 
$\gamma$ varies between $4/3$ and $5/3$, corresponding to relativistic and non-relativistic flows, respectively. The sound speed is 
given through the relation $c^2_s = {\left(1+1/n \right) P}/\rho$. For our study, we vary the 
entire range of adiabatic index gamma from $4/3$ to $5/3$. 

Before proceeding further, let us focus on the velocity-dependent NAP given in TR13. In dimensionless unit, the corresponding generalized potential in Boyer-Lindquist coordinate system is given by  

\begin{eqnarray}
\Phi_{\rm NAP} \left(r, \dot r, \dot \omega \right) = -\frac{1}{r} - \frac{2}{r-2} \, \left[ \left(\frac{r-1}{r-2}\right) {\dot r}^2 \, + \, \frac{r^2 {\dot \Omega}^2}{2} \right], 
\label{1} 
\end{eqnarray}
where, subscript $\rm NAP$ denotes Newtonian analogous potential, and ${\dot \Omega}^2 = {\dot \theta}^2 + \sin^2 \theta {\dot \phi}^2$. The potential is a reduced form of 
\cite{ghosh2016exact} potential in the low energy limit of particle motion. As we are primarily interested in spherical-type flow, neglecting the angular part of the above expression, Eqn. (\ref{1}) reduces to 

\begin{eqnarray}
\Phi_{\rm NAP} \left(r, \dot r \right)= -\frac{1}{r} - \frac{2(r - 1)\dot r^2}{(r-2)^2}.
\label{2}
\end{eqnarray}

One can now compute the corresponding generalized force using the relation 
$\mathscr{F}_{\rm NAP} = - \frac{\partial \phi_{\rm NAP}}{\partial r} + \frac{d}{dt}\frac{\partial \phi_{\rm NAP}}{\partial \dot{r}}$, which leads to

\begin{eqnarray}
\mathscr{F}_{\rm NAP} \left(r, \dot r \right) = - \frac{1}{r^2} + \frac{4(r - 1)}{r^4} + \frac{2\dot r^2}{r(r - 2)}.
\label{3}
\end{eqnarray}
The above expression is identical to the expression given by Eqn. (2.14) in TR13 for purely radial motion. 

Preserving the similar scheme of using existing PNPs in Newtonian hydrodynamical framework, the basic conservation equations for a stationary, inviscid spherical 
accretion flow in the presence of velocity-dependent force $F_{NAP}$ given in Eqn. (\ref{3}) are as follows: \\

(i)  Baryon conservation equation:
\begin{eqnarray}
\frac{d}{dr}\left(4 \pi r^2 \rho v \right) = 0,
\label{4}
\end{eqnarray}

(ii) Radial momentum transfer equation: 
\begin{eqnarray}
v \frac{dv}{dr} + \frac{1}{\rho}\frac{dP}{dr} + \frac{1}{r^2} - \frac{4(r - 1)}{r^4} - \frac{2}{r(r - 2)} v^2= 0
\label{5}
\end{eqnarray}

Here it is to be noted that as a standard practice, in the context of fluid equations, we have identified $\dot r$ as the radial velocity `$v$' of the flow. Integrating Eqn. (\ref{4}), one obtains 
the Baryon mass accretion rate $\dot M$ given by $\dot M = - 4 \pi r^2 \rho v$. Combining Eqns. (\ref{4}) and (\ref{5}) and rearranging, we obtain

\begin{eqnarray}
\frac{dv^2}{dr} = \frac{2\left[\frac{2c^2_s}{r} - \frac{1}{r^2} + \frac{4(r - 1)}{r^4} + \frac{2v^2}{r(r - 2)}\right]}{1 - \frac{c^2_s}{v^2}} = \frac{N1 \, (v, c_s, r)}{D \, (v, c_s)}
\label{6}
\end{eqnarray}
and 
\begin{eqnarray}
\frac{dc^2_s}{dr} = - \frac{1}{n} \frac{c^2_s}{v^2} \frac{\left[\frac{2v^2}{r} - \frac{1}{r^2} + \frac{4(r - 1)}{r^4} + \frac{2v^2}{r(r - 2)}\right]}{1 - \frac{c^2_s}{v^2}} = 
\frac{N2 \, (v, c_s, r)}{D \, (v, c_s)}, 
\label{7}
\end{eqnarray}
respectively, where we identified $`N1 \, (v, c_s, r)'$, $`N2 \, (v, c_s, r)'$ and $`D \, (v, c_s)'$ as numerators and the denominator, respectively.
The dynamics of the accretion flow is described through the Eqns. (\ref{6}) and (\ref{7}). Following the previous authors as before (e.g., \citet{matsumoto1984viscous,1987PASJ...39..309F,chakrabarti1989standing,chakrabarti1990theory,chakrabarti1996grand,lu1997ApLC..35..389L,Das_2002,mukhopadhyay2003global,raychaudhuri2018spherical,raychaudhuri2021multi}), we solve the stated equations employing the standard practice of sonic point analysis. As usual for transonic flows, a smooth and a continuous solution inevitably entails that the following identity $N1 = N2 = D = 0$ must be satisfied 
at a specific radial location, termed as `critical radius' which in our case is the `sonic radius' ($r_c$). At $r_c$, the above identity renders
 
\begin{eqnarray}
\left. v^2 \right\vert_{r_c} \, = \, \left. c^2_s \right\vert_{r_c} \, = \, \frac{1}{2 r_c} \frac{\left(1-2/r_c \right)^3}{\left(1-1/r_c \right)} 
\label{8}    
\end{eqnarray} 
At sonic location $r_c$, we apply l’Hospital’s rule to respective Eqns. (\ref{6}) and (\ref{7}), that yields

\begin{eqnarray}
\left.\frac{d v^2}{dr} \right\vert_{r_c} = \frac{- \, \mathcal{B} \, \pm \, \sqrt{{\mathcal{B}}^2 - 4 \, \mathcal{A} \mathcal{C} }}{2 \mathcal{A} } 
\label{9}
\end{eqnarray}
and 
\begin{eqnarray}
\left.\frac{d c^2_s}{dr} \right\vert_{r_c} = - \frac{1}{N} \left[\mathcal{F}_c  
+ \frac{1}{2} \left.\frac{d v^2}{dr} \right\vert_{r_c} \right] \, , 
\label{10}
\end{eqnarray}
respectively, where 
\begin{equation}
\begin{rcases*}
\mathcal{A} = \frac{1}{2} \, (2 n + 1 )/n, \\
\mathcal{B} = \frac{1}{n} \left(\frac{2 v^2_c}{r_c} + \mathcal{F}_c \right) - \frac{4 v^2_c}{r_c \left(r_c - 2 \right)}, \\
\mathcal{C} = 2 u^2_c \, \left[ \frac{2 c^2_{\rm sc}}{r^2_c} + \frac{2 \mathcal{F}_c}{n r_c} + \left( - \frac{2}{r^3_c} - \frac{4 \left(4 - 3 r_c \right)}{r^5_c} 
+ \frac{4 \left(r_c-1 \right)}{r^2_c \left(r_c -2 \right)^2}  \right) \right],  
\end{rcases*}
\label{11}
\end{equation}
with $\mathcal{F}_c = \frac{1}{r^4_c}\frac{\left(r_c -2 \right)^3}{\left(r_c -1 \right)}$. In Eqn. (9), `$-$' sign in the `$\pm$ indicates accretion solution, and 
the `$+$' sign indicates the wind solution. 

To understand the transonic behavior of the flow, it is necessary for us to compute `sonic energy' and `sonic entropy' 
and analyze their variation with sonic radius. Usually for non dissipative (inviscid) spherical accretion flow, specific energy of the flow is obtained by simply 
integrating Euler's equation. For which case the energy remains conserved throughout the accretion flow regime, 
and the sonic energy ($\mathscr{E}_c$) is then equivalent to energy at outer accretion boundary. For our case, however, the velocity-dependent potential (or force) 
given in Eqn. (\ref{2}) [or in Eqn. (\ref{3})] is not a conserved quantity, and consequently, the energy of the flow will not remain conserved at all radii. Nonetheless, following 
the standard practice used for dissipative accreting system (e.g., \citet{chakrabarti1990theory,chakrabarti1996grand}), one can define the specific energy at sonic location 
`$\mathscr{E}_c$', which for our case is given by 

\begin{eqnarray}
\mathscr{E}_c = \left.{\rm K.E} \right\vert_{r_c} +\left. {\rm T.E} \right\vert_{r_c} + \left.\Phi_{\rm NAP} \left(r, \dot r \right) \right\vert_{r_c} - \left.{\dot r} \frac{d\Phi_{\rm NAP}}{d{\dot r}} \right\vert_{r_c}, 
\label{12}
\end{eqnarray}
where {\rm T.E} is the thermal energy. Identifying $\dot r$ as $v$, the expression for specific energy at $r_c$ then follows 

\begin{eqnarray}
\mathscr{E}_c = \frac{1}{2} v^2_c + n c^2_{\rm sc} - \frac{1}{r_c} + \frac{2 (r_c - 1)}{(r_c - 2)^2} \, v^2_c. 
\label{13}
\end{eqnarray}
Note that, an appropriate value of $\mathscr{E}_c$ needs to be provided to solve the dynamical equations, as a boundary condition for the flow. Using the expressions 
for $v_c$ and $c_{\rm sc}$, and rearranging Eqn. (\ref{13}), one obtains a polynomial equation in $r_c$, given by

\begin{eqnarray}
2\mathscr{E}_c \, r^4_c - \left[2\mathscr{E}_c + n - 3/2 \right] \, r^3_c - (1 -6n) \, r^2_c + 12n\, r_c - 8 n = 0. 
\label{14}
\end{eqnarray}
The specific entropy of the flow is expressed through the relation $\dot {\mathcal{\mu}} =  (\gamma K)^n \vert {\dot M} \vert$ 
(e.g., \citet{chakrabarti1989standing, chakrabarti1990theory, chakrabarti1996grand,raychaudhuri2018spherical,raychaudhuri2021multi}). At sonic location, the expression for sonic entropy is given by 

\begin{eqnarray}
\dot {\mathcal{\mu}}_c =  4 \pi r^2_c \left[\frac{\left(r_c - 2 \right)^3}{2r^3_c \left(r_c - 1 \right)} \right]^{(2n+1)/2}. 
\label{15}
\end{eqnarray}

Classical Bondi-type accretion onto a point gravitating source is a transonic flow usually characterized by a single `saddle-type' sonic point; the physically meaningful 
critical point through which the physical accretion flow passes continuously connecting infinity to BH event horizon, entering the BH supersonically. 
However, it would be interesting to check, whether such {\it{`criticality-condition'}} would remain intact or gets violated in the fluid flow, for our scenario. Here, we examine the 
transonic behavior of adiabatic class of such spherical Bondi-type flows in the presence of velocity-dependent potential [as described in Eqn. (2)] in the standard Newtonian hydrodynamical 
framework we adopted here. One can determine the transonic behavior of fluid flows from Eqn. (\ref{14}) and/or Eqn. (\ref{15}), through $\mathscr{E}_c - r_c$, and/or 
$\dot {\mathcal{\mu}}_c - r_c$ profiles, respectively. 

\begin{figure*}
\centering
\includegraphics[width=150mm]{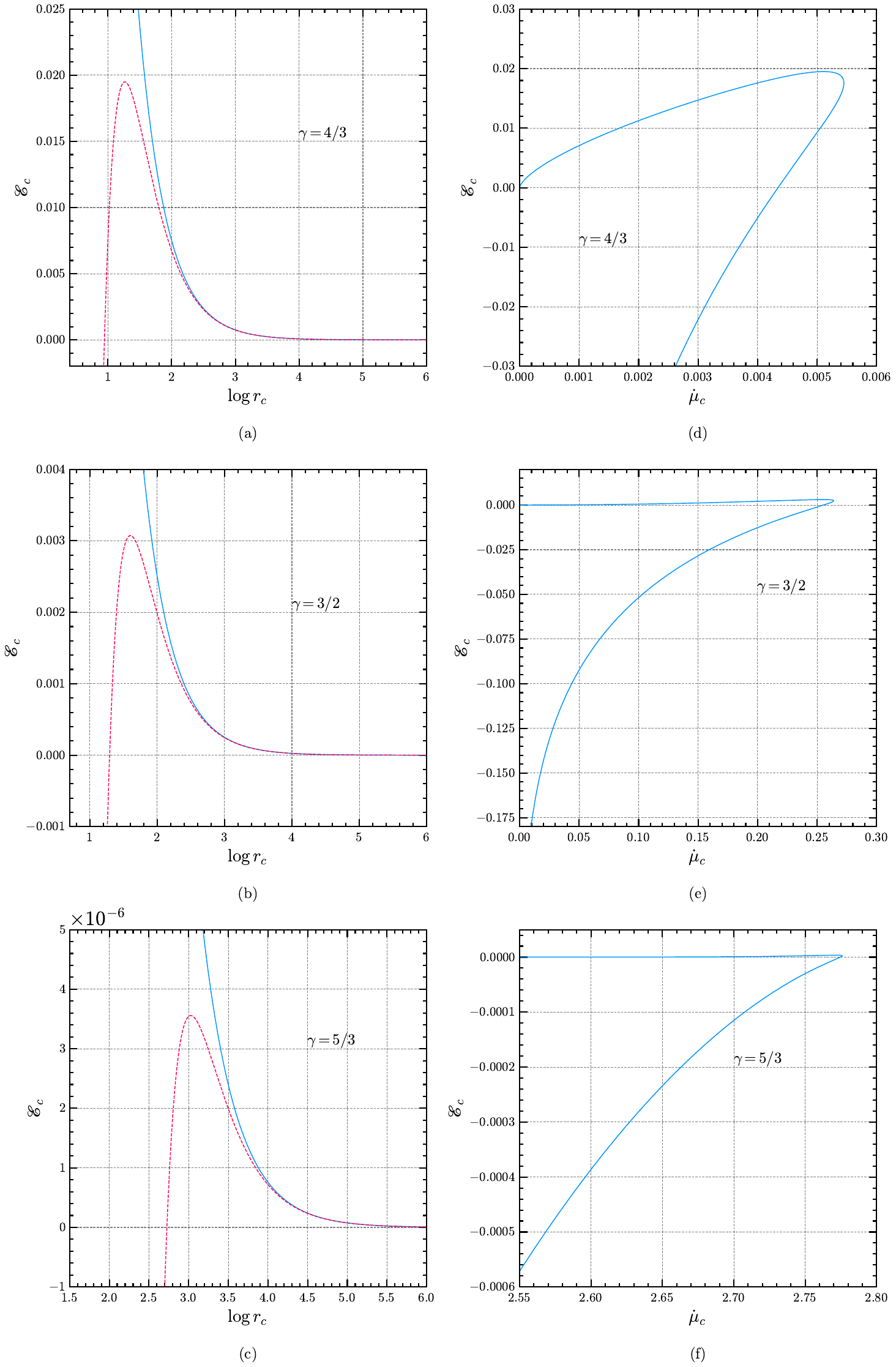}
\caption{ Variation of sonic energy $\mathscr{E}_c$ as functions of sonic location $r_c$ [figures \ref{Fig1}(a),(b),(c)] and sonic entropy $\dot {\mathcal{M}}_c$ [figures
\ref{Fig1}(d),(e),(f)], corresponding to our velocity-dependent potential, for different values of $\gamma$. In curves (\ref{Fig1}(a),(b),(c)), solid ({\it blue}) line corresponds to 
classical Bondi case, where as dotted ({\it red}) line correspond to velocity-dependent potential. In Figs. \ref{Fig1}(d),(e),(f), the curves correspond to velocity-dependent 
potential. Note that $r_c$ is expressed in units of $r_g = G M_{\rm BH}/c^2$. 
 }
\label{Fig1}
\end{figure*}

Figures \ref{Fig1}(a),(b),(c) depict the variation of $\mathscr{E}_c$ as a function of $r_c$ corresponding to different values of 
adiabatic constant $\gamma$. The profiles indicate a departure of the flow from the that of classical Bondi solution, with the likely emergence of {\it{two sonic points}} 
(or two critical points) in the flow at all radii outside the horizon. Let us now focus 
on any one of the $\mathscr{E}_c - r_c$ profiles. The outer sonic points having negative slopes of the curve are all likely to be `saddle-type' 
or `X-type' sonic points through which the flow attains supersonic speed. Conversely, the inner sonic point branch having the positive slope does not likely to be 
consisting of `saddle-types', rather `centre-types' (`O-types') or `spiral-types'. Nonetheless, the outer sonic points almost coincide with that of the sonic locations 
corresponding to classical Bondi case. Even though, in two sonic-point scenario, had both the sonic points 
being `saddle-types', it would not have been possible to connect two sonic points through continuous solutions (e.g., \citet{mandal2007critical}). On the other hand, in the context of 
well-behaved fixed points, it has been pointed out before that two adjacent fixed points can never be both `saddle-types' \citep{jordan2007nonlinear}

In Figs \ref{Fig1}(d),(e),(f) we depict the variation of $\mathscr{E}_c$ as a function of $\dot {\mathcal{\mu}}_c$. The profiles show that the curves
do not (or tend to) form closed loops, as is being expected from the analysis of $\mathscr{E}_c - r_c$ profiles. The upper portion branch of the curves plausibly indicate outer 
`X-type' sonic point branch. Had thee sonic points emerged in our flow, the curves would have formed closed loops with the likely appearance of 
inner `X-type' sonic point branch in $\mathscr{E}_c - \dot {\mathcal{\mu}}_c$ profiles 
(namely, `swallowtail catastrophe', see for e.g., \citet{chakrabarti1989standing,chakrabarti1990theory}). 

To understand the flow behavior, one needs a robust and thorough analysis about the nature of the critical points, which we pursue in the following section. 

\begin{figure*}
\centering
\includegraphics[width=150mm]{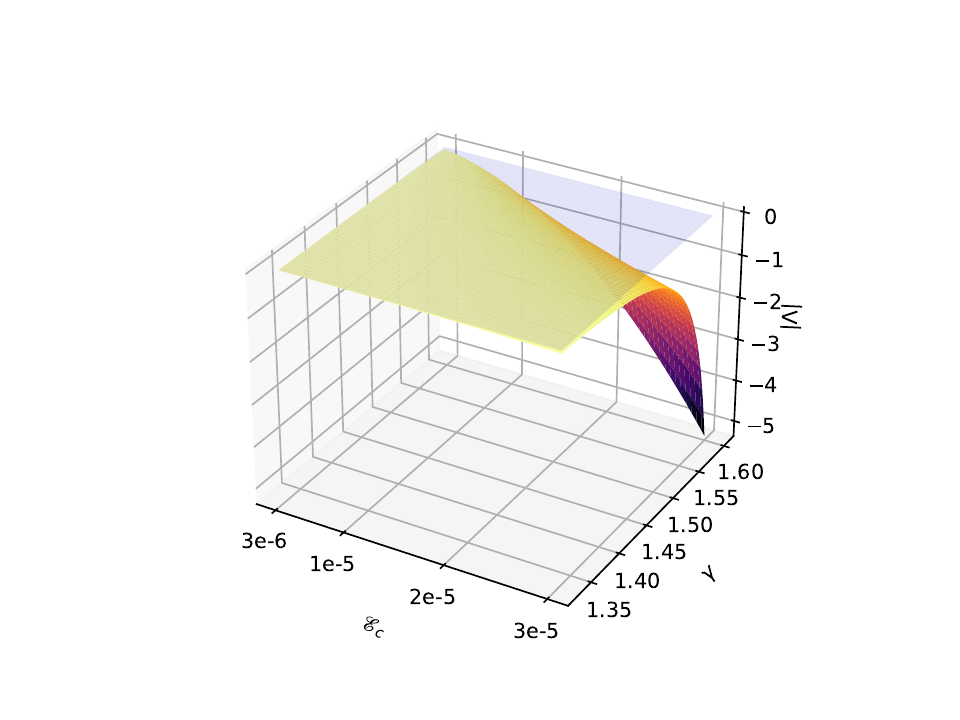}
\caption{ Variation of quantity $|\Lambda|$ (associated with an outer `X-type' sonic point) as a function of parameters 
$\mathscr{E}_c$ and $\gamma$. All values of $|\Lambda|$ are negative for the range of $\mathscr{E}_c$ and $\gamma$ appropriately chosen here. 
 }
\label{Fig2}
\end{figure*}

\begin{figure*}
    \centering
    \begin{subfigure}[b]{0.4\textwidth}
        \centering
        \includegraphics[width=\textwidth]{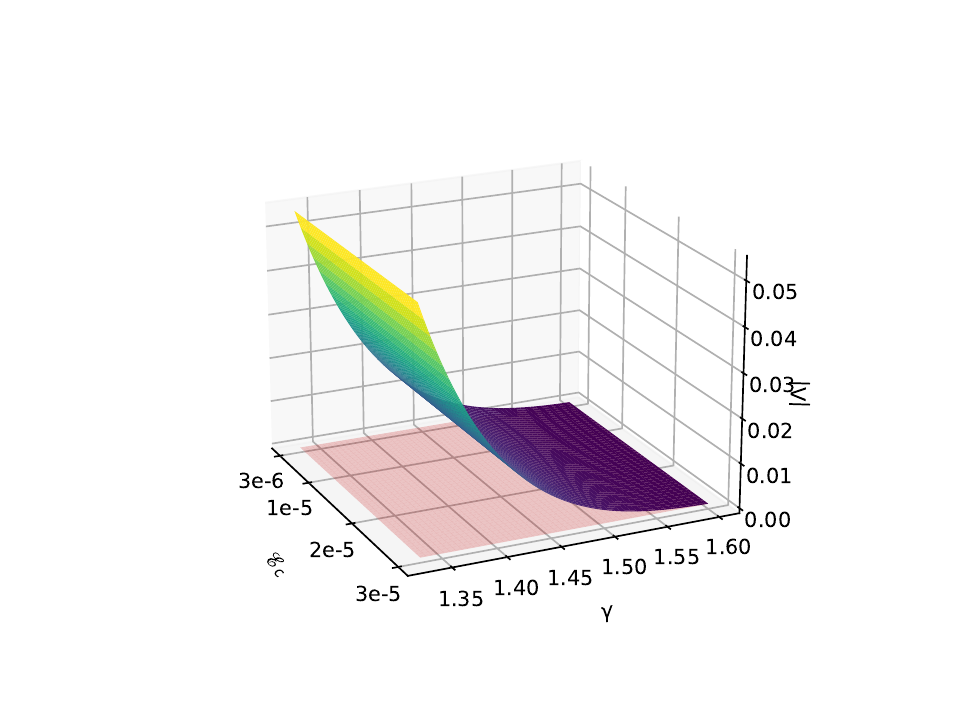}
        \caption{}
        \label{fig:spnt_1}
    \end{subfigure}
    \hfill
    \begin{subfigure}[b]{0.4\textwidth}
        \centering
        \includegraphics[width=\textwidth]{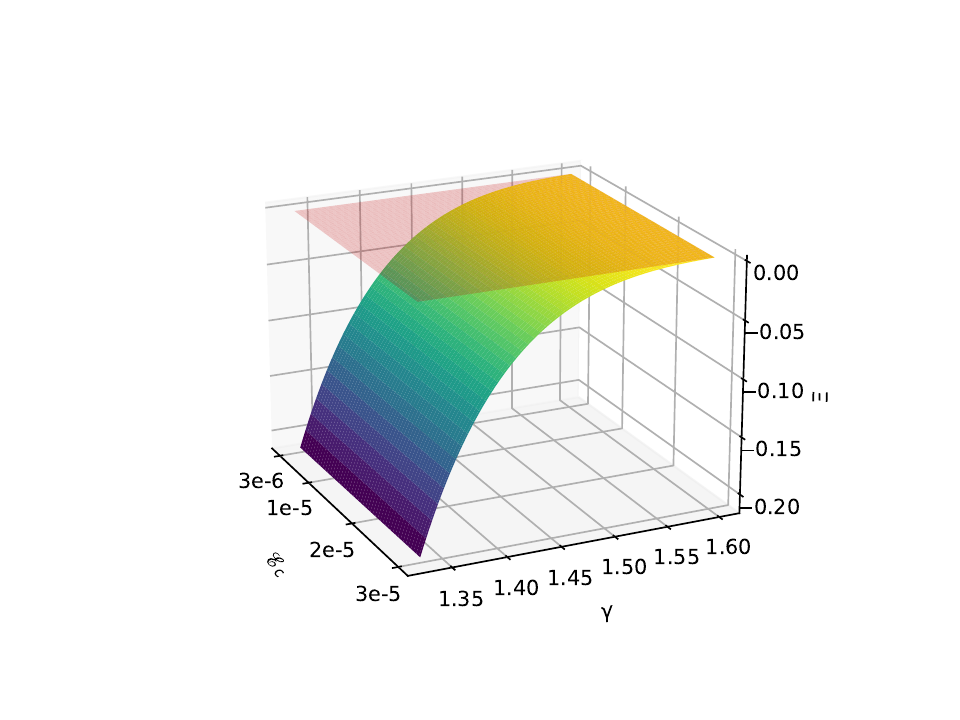}
        \caption{}
        \label{fig:spnt_2}
    \end{subfigure}
    \caption{  Variation of quantity $|\Lambda|$ and $\Xi \equiv Tr(\Lambda)^2 - 4 |\Lambda|$ (associated with an inner `spiral-type' sonic point) as a function 
of parameters $\mathscr{E}_c$ and $\gamma$. All values of $|\Lambda|$ are positive, whereas all values of quantity 
$\Xi$ are negative [as shown in figures (\ref{fig:spnt_1} )and (\ref{fig:spnt_2}), respectively], for the range of $\mathscr{E}_c$ and $\gamma$ chosen here.
  }
    \label{Fig3}
\end{figure*}



\section{Properties of the critical points}  
\label{sec:crit}

Following, for e.g., \citet{mandal2007critical,kato2020fundamentals}(hereinafter KF20); \citet{yang2021spherical}, and references in them, one can represent the steady, spherically symmetric flow as an autonomous dynamical system. Details have been provided in KF20. We do not show it here. Here we follow the procedure as given in KF20. Following KF20, we recast Eqn. (\ref{9}) in the 
following form 

\begin{eqnarray}
\left.\lambda_{12}\left(\frac{dv^2}{dr} \right\vert_{r_c} \right)^2 + \left.(\lambda_{11}-\lambda_{22})\left(\frac{dv^2}{dr} \right\vert_{r_c} \right) + \lambda_{21} = 0,
\label{16}
\end{eqnarray}
where $\lambda$s are given by:    

\begin{eqnarray}
\nonumber \lambda_{11} = \left.\frac{dD}{dr} \right\vert_{r_c}, \; \lambda_{12} = \left.\frac{dN1}{dv^2} \right\vert_{r_c} \\
\lambda_{21} = \left.\frac{dD}{dr} \right\vert_{r_c}, \; \lambda_{22} = \left.\frac{dN1}{dv^2} \right\vert_{r_c},  
\label{17}
\end{eqnarray}
which for our case are  

\begin{equation}
\begin{rcases*}
\lambda_{11} = \frac{2}{n r_c}, \\
\lambda_{12} = \frac{1}{v_c^2} \left(\frac{1}{2n} + 1\right), \\
\lambda_{21} = - \frac{4 v_c^2}{r_c^2}\left(\frac{2}{n} + 1\right) - \frac{8 v_c^2 (r_c - 1)}{r_c^2 (r_c -2)^2} + \frac{4 (r_c -2) (r_c -4)}{r_c^5},  \\
\lambda_{22} = \frac{4}{r_c (r_c - 2)} - \frac{2}{n r_c} 
\end{rcases*}
\label{18}
\end{equation}


We define a Jacobian matrix $\Lambda$ of the dynamical system at the critical point $(r_c, u_c)$ having $\lambda_{ij}$ as its elements, given by 

\begin{equation}
\label{19}
\Lambda = \begin{bmatrix}
\lambda_{11} & \lambda_{12} \\
\lambda_{21} & \lambda_{22} 
\end{bmatrix} 
\end{equation}
where $\lambda^{'s}_{ij}$ are defined in Eqn. ({\ref{18}). 

Depending on the values of determinant of $\Lambda (|\Lambda|)$, the trace of $\Lambda [Tr(\Lambda)]$, and the quantity $\Xi \equiv Tr(\Lambda)^2 - 4 |\Lambda|$, the nature of critical 
points of the dynamical system can be determined (KF20). For the sake of completeness, the conditions are tabulated in Table \ref{tab:nature} as follows:

\begin{table}
\caption{Nature of the critical points}
\label{tab:nature}
\centering
\begin{tabular}{|c|c|}
\hline
Conditions & Types of critical points \\ \hline 
$|\Lambda| <  0 $ & Saddle \\ \hline
$|\Lambda| > 0 $ and $Tr(\Lambda)^2 - 4 |\Lambda| \geq 0 $  & Nodal \\ \hline
$|\Lambda| > 0 $ and $Tr(\Lambda) = 0 $  & Centre \\ \hline
$|\Lambda| > 0 $ and $Tr(\Lambda)^2 - 4 |\Lambda| \leq 0 $  & Spiral\\ \hline 
\end{tabular}
\end{table}

The parameter space in our case is defined by sonic energy $\mathscr{E}_c$ and adiabatic constant $\gamma$. Figure \ref{Fig2} shows that for all relevant values of $\mathscr{E}_c$ 
and $\gamma$ ($4/3 \leq \gamma \leq 5/3$), the corresponding outer sonic points always render $|\Lambda|$ to be negative. On the other hand, corresponding to inner sonic 
points, for all relevant vales of $\mathscr{E}_c$ and $\gamma$, $|\Lambda|$ yields positive values, whereas the corresponding quantity $\Xi \equiv Tr(\Lambda)^2 - 4 |\Lambda|$ always yields 
negative values, as depicted in Fig. \ref{Fig3}. Moreover, it is clearly seen that for all values of $r_c > 2$, $Tr(\Lambda)$ is always non-zero. Figures \ref{Fig2} and \ref{Fig3} clearly demonstrates that for the range of $\mathscr{E}_c$ and $\gamma$ appropriately chosen here, outer sonic point is always a `saddle-type' 
or (`X-type'), whereas, the inner one is throughout `spiral' in nature. 

\begin{figure*}
    \centering
    \begin{subfigure}[b]{0.4\textwidth}
        \centering
        \includegraphics[width=\textwidth]{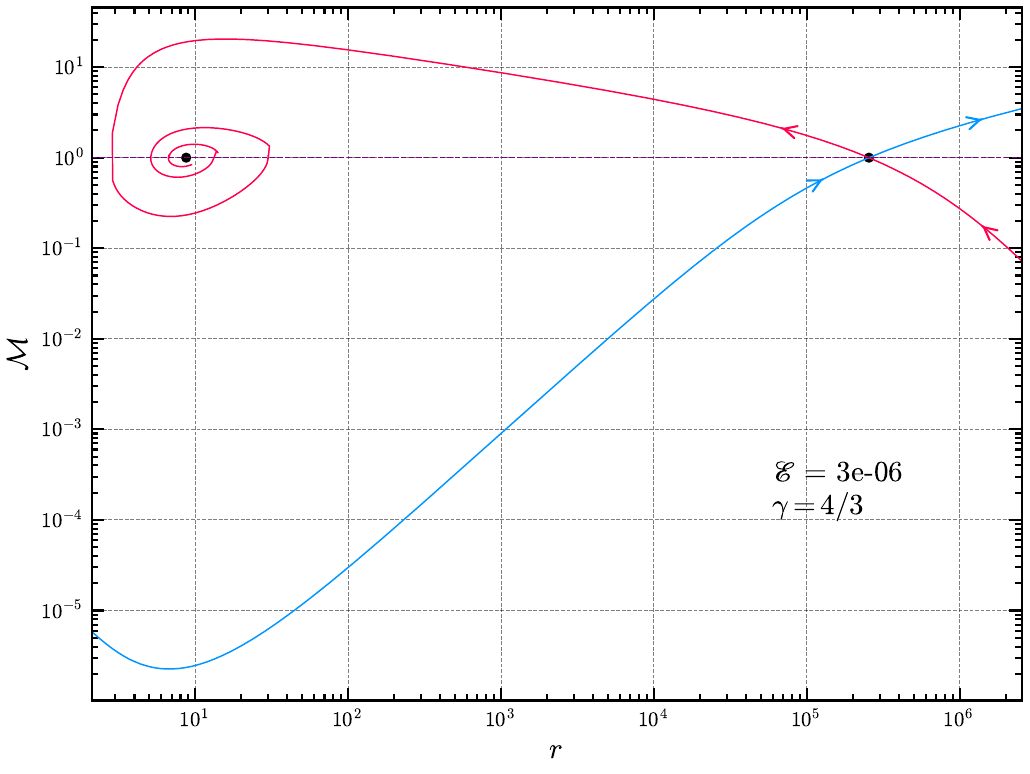}
        \caption{}
        \label{fig:prof133}
    \end{subfigure}
    \hfill
    \begin{subfigure}[b]{0.4\textwidth}
        \centering
        \includegraphics[width=\textwidth]{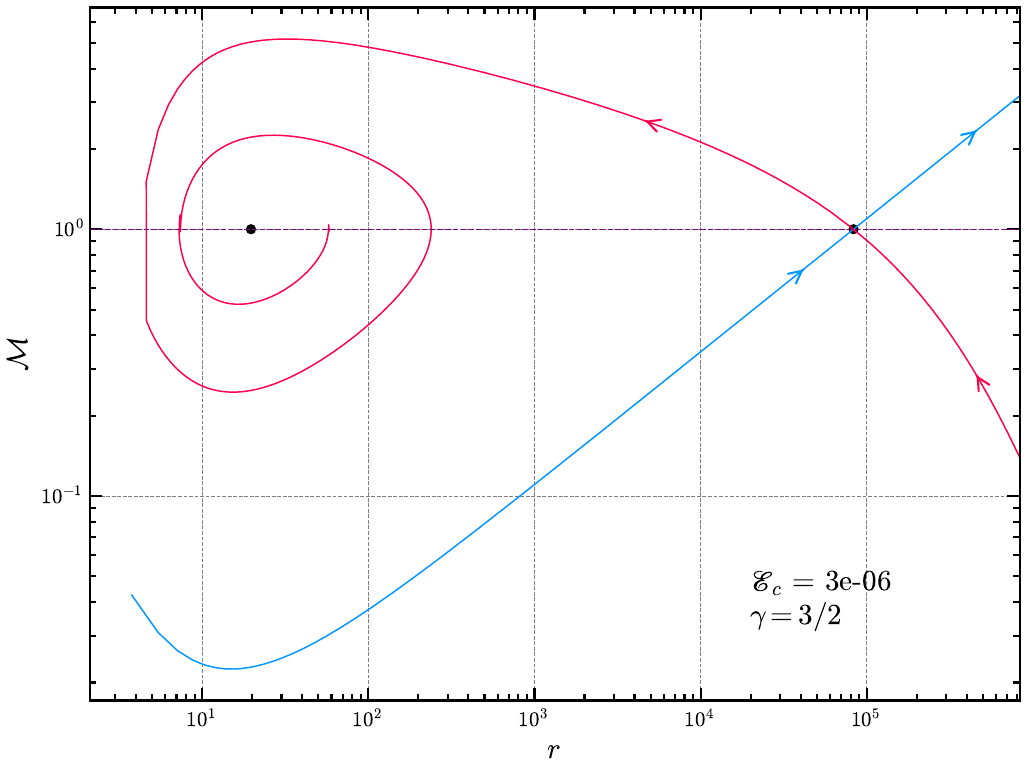}
        \caption{}
        \label{fig:prof150}
    \end{subfigure}
    \caption{ Variation of Mach number $(M)$ as a function of radial distance $(r)$ representing phase portraits for a few sample cases corresponding to ($\mathscr{E}_c$, $\gamma$) pairs  of parameters. The sonic Points are denoted by solid black dots. Outer dot represents outer ``X-type'' sonic point, whereas the inner dot represent inner ``spiral-type'' sonic 
point. Note that $r$ is expressed in units of $r_g = G M_{\rm BH}/c^2$. The corresponding value of $\mathscr{E}_c$ for which the profiles are generated is given in the respective figures.  
 }
    \label{Fig4}
\end{figure*}

\begin{figure*}
    \centering
    \begin{subfigure}[b]{0.24\textwidth}
        \centering
        \includegraphics[width=\textwidth]{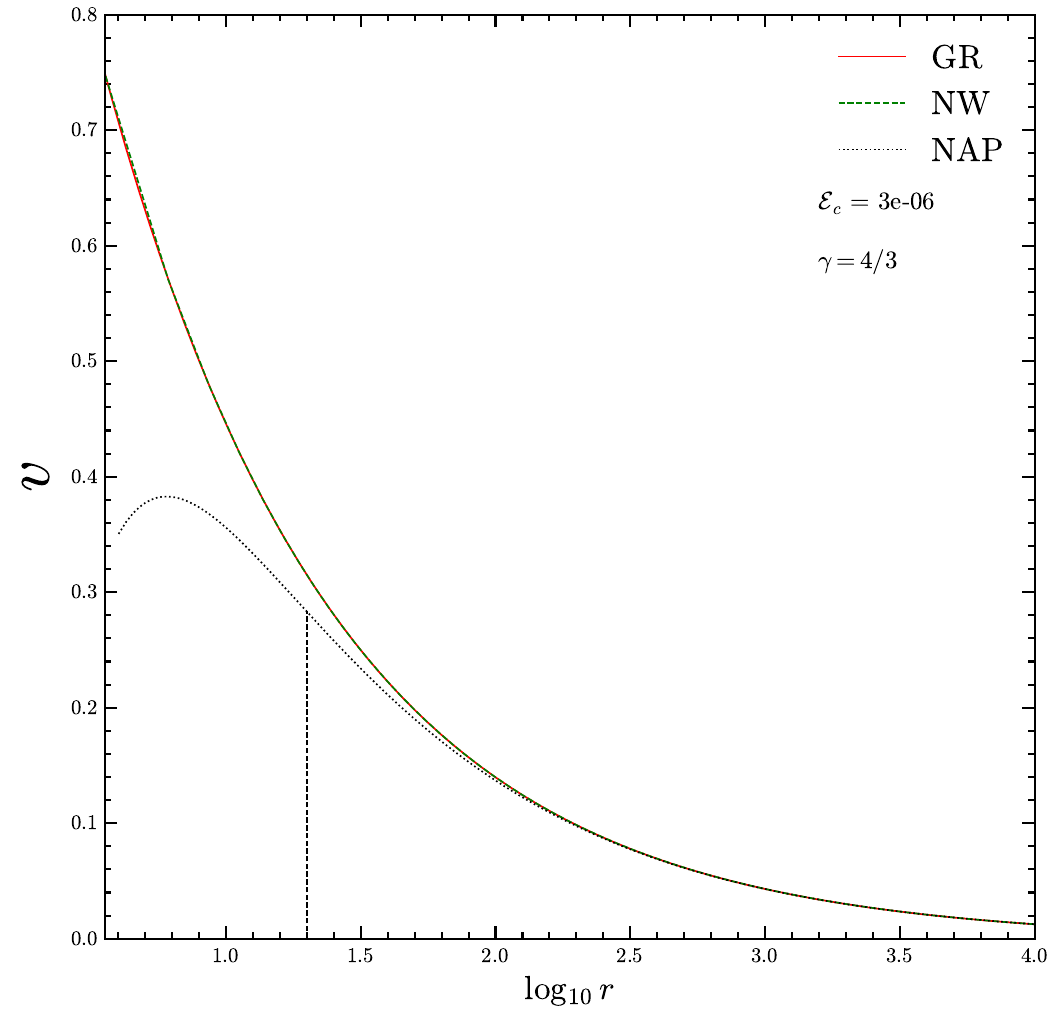}
        \caption{}
        \label{fig:vel133}
    \end{subfigure}
    \hfill
    \begin{subfigure}[b]{0.24\textwidth}
        \centering
        \includegraphics[width=\textwidth]{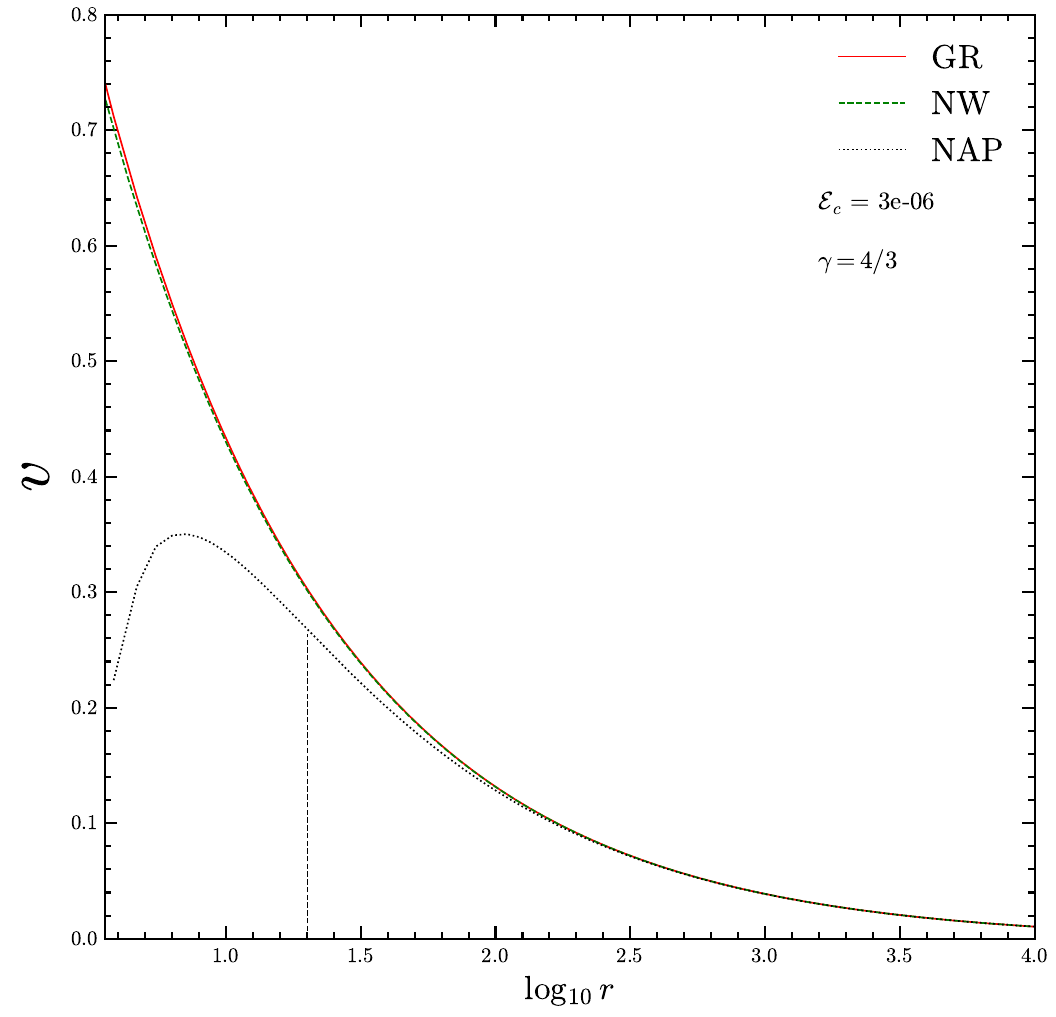}
        \caption{}
        \label{fig:vel150}
    \end{subfigure}
      \hfill
      \begin{subfigure}[b]{0.24\textwidth}
        \centering
        \includegraphics[width=\textwidth]{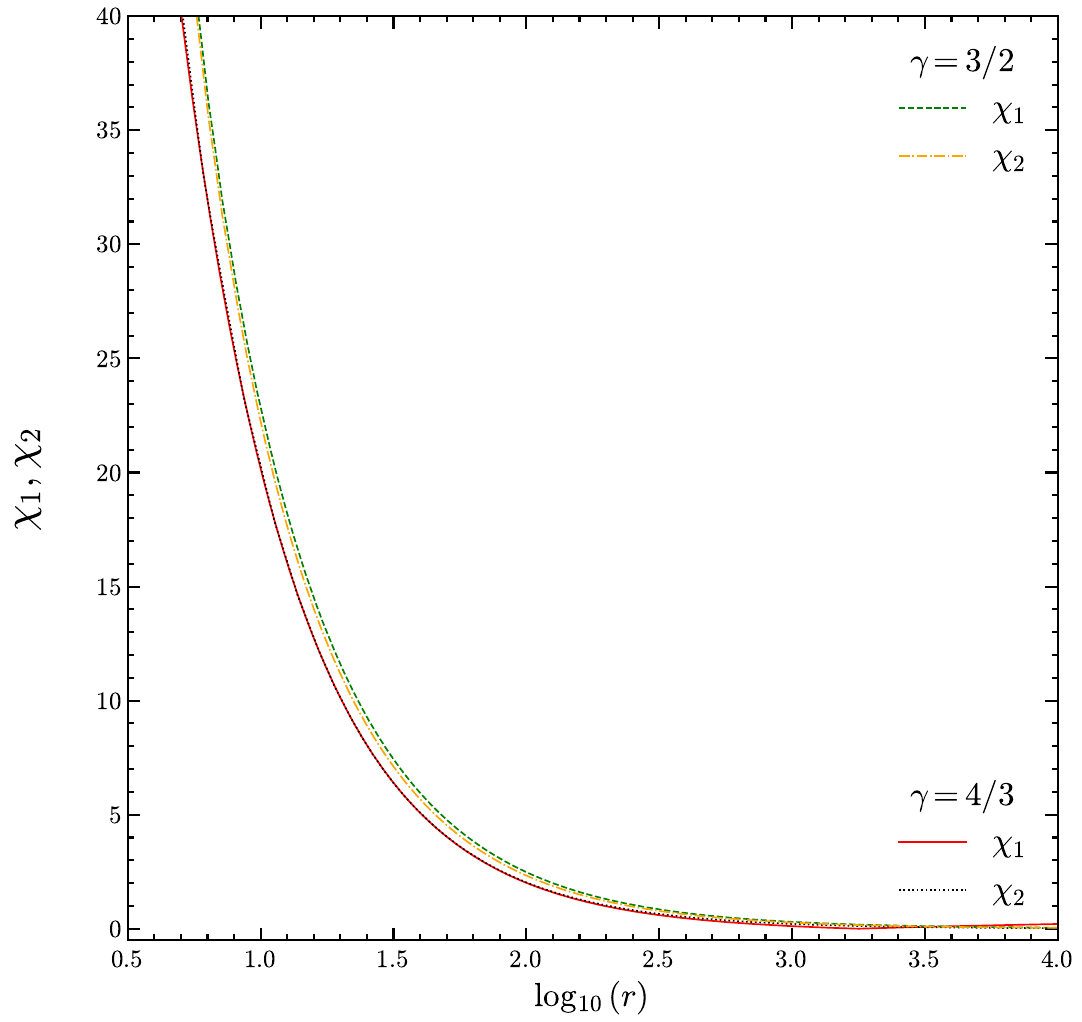}
        \caption{}
        \label{fig:velrel}
    \end{subfigure}
    \caption{ Variation of radial velocity $(v)$ as a function of radial distance $(r)$ for a few sample cases corresponding to ($\mathscr{E}_c$, $\gamma$) pairs of parameters, for the matter
    passing through the outer sonic point, and a comparison with exact GR and Newtonian Bondi (NW) case. 
    In panels (a) and (b), solid ({\it red}) line corresponds to GR case, dashed ({\it green}) line corresponds to Newtonian Bondi (NW) case, where as dotted ({\it black}) line 
    corresponds to NAP. Vertical lines in both the panels denote the radius, where the deviation from exact relativistic value is approximately $ 10 \%$. Note that $r$ is expressed in 
    units of $r_g = G M_{\rm BH}/c^2$. The corresponding value of $\mathscr{E}_c$ for which the profiles are generated is given in the respective figures. In panel (c), we 
    show the relative differences (in percentage) between GR and NAP cases (denoted by quantity $\chi_1$), as well as between Newtonian Bondi (NW) and NAP cases 
    (denoted by quantity $\chi_2$) corresponding to the solutions in panels (a) and (b). In panel (c), solid ({\it green}) and dotted ({\it red}) lines correspond to $\chi_1$ 
    and $\chi_2$, respectively, for $\gamma = 4/3$ case, while, dashed ({\it orange}) and dotted-dashed ({\it black}) lines are similar to solid and dotted lines, respectively, 
    but for $\gamma = 3/2$ case. 
 }
    \label{Fig5}
\end{figure*}



\section{Fluid properties and global topological phase portraits}
\label{sec:topol}

The foregoing sections elucidate that the fluid system of our present interest exhibits two sonic points in the flow. 
Out of two sonic points, the outer sonic point is always ``saddle-type'' (or ``X-type'') through which the sonic transonic 
occurs, with the flow attaining supersonic speed. Conversely, the inner sonic point is found to be ``spiral-type'', which is unphysical, as no stationary solution 
will traverse through it. To obtain the transonic flow profiles and a family of solutions, one needs to supply the value of $\mathscr{E}_c$ as the flow boundary 
condition at outer sonic point. For a given value of $\mathscr{E}_c$, using Eqn. (\ref{14}), one would then be able to determine the corresponding value of $r_c$. 
Subsequently, one could then evaluate the fluid velocity and sound speed at $r_c$ using Eqn. (\ref{8}). These would then be supplied as additional boundary conditions of 
the flow to evaluate $v$ and $c_s$ from Eqns. (\ref{6}) and (\ref{7}). Employing fourth-order Runge–kutta method, the said equations are then solved, integrating from the outer 
sonic point inwardly and outwardly, simultaneously. 

In Fig. \ref{Fig4} we show the flow topologies or phase portraits for a few sample cases, in the context of the fluid system of our interest, depicting 
outer ``X-type'' sonic point, and inner ``spiral-type'' sonic point. Here it is important to note that in the context of matter falling into the BH, any physical transonic accretion flow should pass through ``X-type'' sonic point before entering BH horizon supersonically. Although for our case, one does have an outer ``X-type'' sonic point, but the solution will not extend supersonically till the horizon, as the inner sonic point appearing to be unphysical (``spiral-type''), will not allow any stationary solution to pass through it. Figure \ref{Fig5}(a),(b) shows the variation of 
radial velocity $(v)$ as a function of radial distance $(r)$, for the matter passing through the outer ``X-type'' sonic point, and a comparison with the exact GR and Newtonian 
Bondi cases. Here, relevant for our purpose, in the context of GR solution, we have appropriately chosen the radial component of fluid's ``three-velocity'' as measured by a local, stationary (static) observer [see e.g., \citet{shapiroteukolsky83,aguayo2021spherical}], following \citet{mandal2007critical}, as we make a comparison with respect to fluid velocity in 
the ``Newtonian framework''. This ``three-velocity'' is related to the radial component of fluid's ``four-velocity ($u$)'' [as used in \citet{michel1972accretion}] through the following transformation 
relation (in dimensionless unit) [e.g., Appendix G in \citet{shapiroteukolsky83}]
\begin{eqnarray}
v =\frac{u}{(1-2/r + u^2)^{1/2}} \, . 
\label{20}
\end{eqnarray}
Also, in terms of this fluid ``three-velocity'', it follows that at $r = r_c$, $v = c_s$, that actually justifies calling 
$r_c$ as sonic radius or sonic point \citet{shapiroteukolsky83,aguayo2021spherical,mandal2007critical}. This is similar to the scenario in the ``Newtonian framework'' (whether in context of Newtonian potential or TR13 NAP). 
a choice of a comparatively higher value of
the energy, or even a lower value, will render an unrealistic
value of the ambient temperature.

From the figure, we observe that around $\sim 10$ Schwarzschild radii, radial velocity $v$ deviates by about $\sim 10\%$ of the exact relativistic value 
corresponding to $\gamma \sim 4/3$. Similarly for $\gamma \sim 3/2$, a $\sim 10\%$ deviation of $v$ from exact relativistic value occurs around $\sim 11$ Schwarzschild radii. The actual deviation of radial velocity from classical Bondi case however occurs at around $\sim (40 - 50)$ Schwarzschild radius. For higher values of $\gamma$ as $\gamma$ approaches $5/3$, the deviation in $v$ will 
occur from much larger radii, with higher error in $v$ around $10$ Schwarzschild radius. In Fig. \ref{Fig5}(c), for more clarity, we show the relative differences between GR and 
NAP cases, as well as between Newtonian Bondi (NW) and NAP cases, corresponding to the solutions in panels (a) and (b). The (percentage) relative deviation is denoted by the quantity $\chi \, \left(\chi_1 = \frac{\big\vert \, v \vert_{\rm GR} - v \vert_{\rm NAP} \, \big\vert}{v \vert_{\rm GR}} \times 100; \chi_2 = \frac{\big\vert \, v \vert_{\rm NW} - v \vert_{\rm NAP} \, \big\vert}{v \vert_{\rm NW}} \times 100  \right)$, as shown in Fig. \ref{Fig5}(c) }. In both figures \ref{Fig4} and \ref{Fig5}, we choose the value of 
energy $\mathscr{E}_c = 3 \times 10^{-6}$ as boundary condition of the flow. This choice of the value of energy roughly corresponds to a temperature $T$ in the range 
$\left(\sim 6 \times 10^6 - 10^7 \right) {\rm K} $ for $4/3 \lesssim \gamma \lesssim 5/3$, a reasonable scenario for the interstellar medium (ISM) at the center of a galaxy 
[e.g., \citet{narayan2011bondi,raychaudhuri2018spherical}]. For a slight increase in the value of energy $\mathscr{E}_c$, there will be no 
appreciable change in the radial-limit, mentioned before. On the other hand, for a decrease in $\mathscr{E}_c$, the deviation in $v$ from that of classical Bondi case will occur 
from a slightly lesser radius. However, note that, a choice of comparatively higher value of energy, or even a lower value, will render unrealistic value of ambient 
temperature.   

It is interesting to note that, in the context of GR case, at large $r$, away from the horizon, fluid's ``three-velocity ($v$)'' $\simeq$ ``four-velocity ($u$)'' [as 
seen from the Eqn. (\ref{20}); also see \citet{shapiroteukolsky83}]. Again at $r << r_c$, as the flow moves inward, square of the ``four-velocity'' ``$u^2 \approx 2/r$'' 
[see e.g., Appendix G in \citet{shapiroteukolsky83} (Eqn. G.34)], implying that ``$v \simeq u$'' [from Eqn. (\ref{20})]. Thus in the context of GR case, the solution in terms of radial 
component of ``three-velocity'' (that we appropriately used in our study) will remain almost identical to that of solution in terms of `four-velocity'. We have verified this by actually finding the velocity solution in terms of ``four-velocity'', employing the above velocity transformation. We do not show it here. Thus, no significant difference would arise in our findings, had radial component 
of ``four-velocity'' is instead being used, for comparison. 

Here, it is worthy to mention that, had three sonic points (or critical points) appeared in our flow [as being found in advective type accretion flows (see Chakrabarti and collaborators) or 
in case of Bondi-type flows [e.g., \citet{raychaudhuri2018spherical,raychaudhuri2021multi}], and where both inner and outer sonic points 
being ``saddle-types' (or ``X-types''), the flow would have possibly passed smoothly through either of the ``X-type'' sonic points or pass through the 
inner ``X-type'' sonic point after encountering a shock, reaching BH horizon supersonically, thereby physically connecting infinity to the horizon. 
In the next section, we discuss the plausible reasons behind the discrepancy found in our study.    

We further compute Bondi mass accretion rate ($\dot M_B$) which can be defined in terms of the transonic quantities of the flow, and compare that with the GR case. 
In the steady state, corresponding to our NAP, one can write Bondi accretion rate ($\dot M_B$) in the Newtonian framework, given by 

\begin{eqnarray}
\vert \dot M_{B} \vert_{\rm NAP} = 4 \pi r^2_c \, c_{\rm sc} \, \rho_c \, ,   
\label{21}
\end{eqnarray}
where $\rho_c$ is the density at the (outer) sonic location. Using the relation $\rho_c = \rho_{\rm out} \, \left(\frac{c_{\rm sc}}{c_{\rm out}}\right)^{2/(\gamma -1)}$, 
where $\rho_{\rm out}$ and $c_{\rm out}$ are the corresponding density and sound speed at the outer accretion boundary $r_{\rm out}$ [see \S 6 in \citet{raychaudhuri2018spherical}], and 
using Eqn. (\ref{8}), one can define $\vert \dot M_{B} \vert_{\rm NAP}$ 
in terms of only sonic radius $(r_c)$ and outer boundary conditions. Note that, $\rho_{\rm out}$ 
is effectively the density of the ambient medium. Similarly, for GR case, following \citet{mandal2007critical} 
[see Eqn. (7) in \citet{mandal2007critical}], we make use of the appropriate expression for Bondi accretion rate, in terms of fluid's ``three-velocity'' (see the discussion in 
the previous paragraphs of this section), given as
\begin{eqnarray}
\vert \dot M_{B} \vert_{\rm GR} = 4 \pi r^2_c \, v_c \, \rho_c \, \sqrt{\frac{1 - 2 r^{-1}_c}{1 - v^2_c}}  \, .
\label{22}
\end{eqnarray}
One can obtain the above expression from Eqn. (8) in \citet{michel1972accretion} by replacing `$u$' through `$v$' using Eqn. (\ref{20}). Here we note that, in the expression of 
\citet{mandal2007critical} [Eqn. (7)], the factor of `$2$' arising on the right hand side of the numerator of Eqn. (\ref{22}), is absent, as the authors in their work have expressed 
radius $r$ in units of $2GM/c^2$, while we in the present article have expressed  $r$ in units of $GM/c^2$. 
Using Eqns. (6) and (9) from \citet{mandal2007critical}, one can express $\vert \dot M_{B} \vert_{\rm GR}$ solely in terms of $r_c$ and outer boundary conditions. For 
the case of nonrelativistic baryons accreting from ambient medium (e.g., ISM), as usual one can express $c_{\rm out}$ in terms of $T_{\rm out}$, the temperature of the 
ambient medium, given through $c^2_{\rm out} = {\gamma k_B T_{\rm out}}/{\mu m_p}$, where $k_B$ is the Boltzmann constant, $m_p$, the usual mass of 
proton, and $\mu = 0.592$ is the mean molecular weight for the galactic abundance of Hydrogen and Helium with Hydrogen mass fraction $X = 0.75$. One can then 
estimate the quantity ${\vert \dot M_{B} \vert_{\rm NAP}}/{\vert \dot M_{B} \vert_{\rm GR}}$ in terms of respective sonic locations, and ambient 
temperature $T_{\rm out}$, given through 

\begin{eqnarray}
{\vert \dot M_{B} \vert_{\rm NAP}}/{\vert \dot M_{B} \vert_{\rm GR}} = \frac{\left(\frac{1}{2} \right)^{\frac{\gamma+1}{2(\gamma-1)}} \, r^{\frac{3\gamma-5}{2(\gamma-1)}}_c \, 
\left[\frac{\left(1-2/r_c \right)^3}{\left(1-1/r_c \right)}  \right]^{\frac{\gamma +1}{2(\gamma-1)}} \, \, \, \,   \bigg\vert_{\rm NAP}}{ r^2_c 
\left(\frac{1}{4r_c - 1} \right)^{\frac{\gamma+1}{2(\gamma-1)}} \, \left[\frac{1 - 2 r^{-1}_c}{1 - \left(4r_c - 1 \right)^{-1}} \right]^{1/2} \, \left[\frac{\gamma - 1 \, - \, 1.794 \times 10^{-13} \, \gamma \, T_{\rm out}}{\gamma -1 -\left(4r_c - 1 \right)^{-1} }    \right]^{\frac{1}{\gamma -1 }}    \, \, \, \,  \bigg\vert_{\rm GR}}   \, .  
\label{23}
\end{eqnarray}
Here, $r_c$ in the numerator of Eqn. (\ref{23}) corresponds to (outer) sonic location for NAP, whereas in the denominator it corresponds to GR case. 

For any realistic value of the temperature of ambient medium $T_{\rm out}$, the term $1.794 \times 10^{-13} \,\gamma \, T_{\rm out}$ in the denominator of Eqn. (\ref{23}) is much less 
than the preceding term $\gamma - 1$, and hence can be neglected. It is being found that corresponding to different values of $\gamma$, the quantity 
${\vert \dot M_{B} \vert_{\rm NAP}}/{\vert \dot M_{B} \vert_{\rm GR}}$, i.e., the ratio of Bondi accretion rate corresponding to NAP to that corresponding to GR case, is almost equal to 1.



\section{Discussion and remarks} 
\label{sec:discus}

The advantage of the velocity-dependent NAPs is that they can accurately describe the relativistic features of corresponding spacetime 
geometries, including classical predictable GR effects (see for e.g., \citet{ghosh2016exact}). This is unlike the case for existing PNPs which are primarily 
emphasized to reproduce the innermost stable and bound circular orbits of particle motion. 
Newtonian analogous constructs, on the other hand, are done in a manner intended to correctly reproduce geodesic equations of motion of the 
parent metric. The corresponding velocity-dependent potentials are, thus, more accurate representations of corresponding 
spacetime geometries in analogous Newtonian framework, which are uniquely defined. 
In the context of astrophysical phenomena, where one primarily deals with orbital trajectories or dynamics of particle orbits, like orbital 
precession, chaotic dynamics of orbits, N-body dynamics, or in the context of tidal disruption of stars by 
massive BHs, these velocity-dependent NAPs are extremely useful, and in few cases had already been employed (e.g., TR13; \citet{bonnerot2016disc}). One of the primary aims of the Newtonian 
analogous constructs is to accurately study complex relativistic accretion phenomena around BHs/compact objects in Newtonian framework, 
bypassing GR-hydrodynamics (GRHDs)/GR-magnetohydrodynamics (GRMHDs). It is thus worthwhile to examine, how far one can make use of this formalism in hydrodynamical accretion studies within the framework of standard Newtonian hydrodynamics. The NAP modeled by TR13, under the low-energy limit condition for Schwarzschild spacetime, has been applied to the standard geometrically thin accretion disk and accretion infall of non-interacting particles on to a Schwarzschild black hole, and is found to be in good agreement with the exact relativistic solutions. 

In this article, as a further application, we seek to explore the extent to which TR13 NAP could describe a transonic radial/quasi-radial type (hydrodynamical) accretion flow in Schwarzschild 
spacetime, i.e., Bondi-type hydrodynamical spherical accretion within the framework of standard Newtonian hydrodynamics, by conducting an actual hydrodynamical study of 
accretion flow dynamics. We find that the fluid system of our present interest shows a deviation from classical Bondi scenario from a relatively large outer 
radii, with deviation in radial velocity $v$ being $\sim 10\%$ of the exact relativistic value around $10$ Schwarzschild radius, corresponding to $4/3 \lesssim \gamma \lesssim 3/2$, for 
appropriate boundary condition (as depicted in Fig. \ref{Fig5}). For more higher values of $\gamma$ as $\gamma$ approaches $5/3$, the error in $u$ will be higher. 
Interestingly, as the flow moves inward, a significant departure from classical Bondi solution occurs; instead of obtaining single ``saddle-type'' sonic 
transition in the flow (what is being expected in case of an inviscid spherical accretion flow), two sonic points emerge, with the inner one being 
an (unphysical) ``spiral-type'', and outer being usual ``saddle-type'' (or ``X-type'') sonic point. This indicates that, although the matter will pass through the outer ``X-type'' point, with the flow attaining supersonic speed, however, the solution will not extend supersonically till the BH horizon, as the inner ``spiral-type'' sonic point will not 
allow any stationary solution to pass through it. The position of the outer sonic point, although matches closely to that of the sonic point corresponding to classical Bondi case, or for 
that matter, GR Bondi case. We estimated the Bondi mass accretion rate corresponding to the TR13 NAP, which is found to be almost identical to that of the GR case. 

One can raise a very pertinent point: why does a transonic
radial or quasi-radial hydrodynamical accretion flow towards a
BH in the presence of velocity-dependent TR13 like NAP,
within the standard Newtonian hydrodynamical framework,
shows significant departure from the classical Bondi solution,
even though such NAP can accurately capture most relativistic
features of Schwarzschild spacetime, as well as describe
geometrically thin accretion disks, and reproduce the exact
relativistic solutions for particle streamlines much better than
existing PNPs?. A deeper look into this issue is warranted.
Newtonian analogs of corresponding spacetime geometries
have been constructed relative to distant stationary observers.
In the GR framework, from the point of view of a distant
stationary observer, matter will never appear to reach the BH
horizon in the Schwarzschild spacetime. Our findings in this
regard, that no transonic accretion flow solution is allowed
connecting infinity to the BH horizon, appears to be consistent
with the GR description. To have a scenario of transonic
accretion flow to the horizon, as in GR, it is important to
formulate the problem in the perspective of a local stationary
observer. Here one can raise a relevant question—whether it
would be correct to use the velocity-dependent NAP along with
nonmodified Newtonian hydrodynamics. In this context, \citet{witzany2017pseudo} previously endeavored
to address this aspect (although without undertaking any actual
hydrodynamical study) in the language of ``time reparameterization'' (related to lab frame time and proper time). The
authors highlighted the need for modification of the usual
(standard) Newtonian fluid equations to effectively implement
such a velocity-dependent potential in the Newtonian frame-
work. They concluded that a naive implementation of the
corresponding velocity-dependent potential or force term into
nonmodified Newtonian hydrodynamics would be erroneous
and ``could lead to pathological behavior of the fluid near the
BH horizon.'' In their study, \citet{witzany2017pseudo} attempted to develop a modified hydrodynamical
formalism to implement such velocity-dependent potential,
albeit through an ad hoc approach, in the framework of so-
called ``time reparameterization.'' Nonetheless, it would be
important to note that these aspects related to a ``distant
stationary observer,'' or so-called ``time reparameterization,'' will, perhaps, be more relevant very 
close to the BH horizon. 

A more pertinent issue in this regard is the following: The relativistic Euler equation involves the pressure gradient term coupled with the metric tensor and fluid velocity \citet{shapiroteukolsky83}. In the non-relativistic limit (nrl) (i.e., for $v^2/c^2 << 1$ and $2r_g/r << 1$), the GR Euler equation exactly reduces to the original (Newtonian) Euler 
equation. On the other hand, near the BH horizon, $v^2 \sim \frac{2}{r}$, and thus the pressure gradient term of the GR Euler equation effectively reduces to the corresponding term in the standard (Newtonian) Euler equation. Hence, the Newtonian Bondi solutions remain consistent with the corresponding GR results. In contrast, the Euler-type equation in the presence of velocity-dependent TR13 NAP in the standard Newtonian hydrodynamical framework comprises of separate terms (without any coupling to the pressure gradient) for fluid velocity and mass, without having any offsetting 
terms [see Eqn. (\ref{5})]. Thus the Euler-type equation in the presence of velocity-dependent TR13 NAP (that has been derived directly from spacetime metric), in the framework of standard Newtonian hydrodynamics, does not appear to remain consistent with the corresponding GR hydrodynamics, regardless of the fluid's energy. This might be the primary reason behind the discrepancy 
found in our study. 

Our study suggests that a (modified) hydrodynamical
formalism is needed to effectively implement such a velocity-
dependent NAP to describe transonic spherical or quasi-
spherical accretion flows onto a BH, as well as in other similar
studies. This (modified) formalism should align with the spirit
of TR13 like NAP, while remaining consistent with relativistic
hydrodynamics. A modified hydrodynamical or magnetohy-
drodynamical framework consistent with such a velocity-
dependent potential would be extremely useful in the context of
relativistic accretion flow studies, which could then essentially
circumvent GRHD or GRMHD equations. This is a thorough
exercise that we endeavor to pursue in the near future. Ideally,
any proper hydrodynamical framework should be developed
starting from a kinetic approach. In the context of our velocity-
dependent potential, whether such a framework could be
consistently developed requires thorough examination. We
leave this as an exercise for future work


\section*{Acknowledgments}
The authors are thankful to the anonymous reviewer for his/her insightful questions and critical comments that helped us to improve the manuscript. 
Shubhrangshu Ghosh (SG) is grateful to Harish-Chandra Research Institute (HRI) for their hospitality, where the main part of the present work was carried out. 
SG acknowledge IUCAA, Pune, for their support through the `Visiting Associateship Program'.


\end{document}